\documentclass[%
reprint,
superscriptaddress,
amsmath,
amssymb,
aps,
nofootinbib,
]{revtex4-2}

\usepackage{graphicx}
\usepackage{dcolumn}
\usepackage[export]{adjustbox}
\usepackage{bm}
\usepackage[caption=false]{subfig}

\usepackage{xcolor}
\usepackage{soul}

\usepackage{xr-hyper}

\begin{document}

\title{Novel structures of Gallenene intercalated in epitaxial Graphene}

\author{Emanuele Pompei\textsuperscript{*}}
\affiliation{NEST, Istituto Nanoscienze-CNR and Scuola Normale Superiore, Piazza S. Silvestro 12, 56127, Pisa, Italy}

\author{Katarzyna Skibińska\textsuperscript{\dag}}
\affiliation{NEST, Istituto Nanoscienze-CNR and Scuola Normale Superiore, Piazza S. Silvestro 12, 56127, Pisa, Italy}

\author{Giulio Senesi}
\affiliation{NEST, Istituto Nanoscienze-CNR and Scuola Normale Superiore, Piazza S. Silvestro 12, 56127, Pisa, Italy}

\author{Ylea Vlamidis}
\affiliation{NEST, Istituto Nanoscienze-CNR and Scuola Normale Superiore, Piazza S. Silvestro 12, 56127, Pisa, Italy}
\affiliation{Department of Physical Science, Earth, and Environment, University of Siena, Via Roma 56, 53100, Siena, Italy}

\author{Antonio Rossi}
\affiliation{Center for Nanotechnology Innovation@NEST, Istituto Italiano di Tecnologia, Piazza S. Silvestro 12, 56127, Pisa,
Italy}

\author{Stiven Forti}
\affiliation{Center for Nanotechnology Innovation@NEST, Istituto Italiano di Tecnologia, Piazza S. Silvestro 12, 56127, Pisa,
Italy}

\author{Camilla Coletti}
\affiliation{Center for Nanotechnology Innovation@NEST, Istituto Italiano di Tecnologia, Piazza S. Silvestro 12, 56127, Pisa,
Italy}

\author{Fabio Beltram}
\affiliation{NEST, Istituto Nanoscienze-CNR and Scuola Normale Superiore, Piazza S. Silvestro 12, 56127, Pisa, Italy}

\author{Lucia Sorba}
\affiliation{NEST, Istituto Nanoscienze-CNR and Scuola Normale Superiore, Piazza S. Silvestro 12, 56127, Pisa, Italy}

\author{Stefan Heun}
\affiliation{NEST, Istituto Nanoscienze-CNR and Scuola Normale Superiore, Piazza S. Silvestro 12, 56127, Pisa, Italy}

\author{Stefano Veronesi\textsuperscript{*}}
\affiliation{NEST, Istituto Nanoscienze-CNR and Scuola Normale Superiore, Piazza S. Silvestro 12, 56127, Pisa, Italy}

\date{\today}

\begin{abstract}

The creation of atomically thin layers of non-exfoliable materials remains a crucial challenge, requiring the development of innovative techniques. Here, confinement epitaxy is exploited to realize two-dimensional gallium via intercalation in epitaxial graphene grown on silicon carbide.
Novel superstructures arising from the interaction of gallenene (a monolayer of gallium) with graphene and the silicon carbide substrate are investigated. The coexistence of different gallenene phases, including b010-gallenene and the elusive high-pressure Ga(III) phase, is identified.
This work sheds new light on the formation of two-dimensional gallium and provides a platform for investigating the exotic electronic and optical properties of confined gallenene.
\end{abstract} 

\maketitle

\section{Introduction}\label{sec:intro}

\setlength{\skip\footins}{0.5cm}
\footnotetext[1]{Corresponding authors: E.P.: emanuele.pompei@sns.it \\ \textcolor{white}{space space space space} S.V.: stefano.veronesi@cnr.it}
\footnotetext[2]{Currently at: Faculty of Non-Ferrous Metals, AGH University of Krakow, al. Adama Mickiewicza 30, 30-059, Krakow, Poland}

Recently, various two-dimensional (2D) materials have been realized through confinement at the interface between epitaxial graphene (EG) and silicon carbide (SiC) substrates \cite{AlBalushi2016, Forti2020, Briggs2020, Kim2020, Wundrack2021}. The confinement is achieved via the intercalation of atoms under the EG sheet. This method allows for the fabrication of atomically-thin, ordered 2D materials over large area. Furthermore, the graphene sheet protects the as-produced 2D materials from the environment, making them air-stable \cite{Bersch2019, Dong2024}.
The most commonly employed technique is Confinement Heteroepitaxy (CHet) \cite{Briggs2020}. It involves the deposition of pure elements on the pre-heated EG/SiC substrate in a tube furnace near to atmospheric pressure, followed by thermally activated intercalation. CHet was implemented for the first time for the fabrication of 2D gallium nitride \cite{AlBalushi2016} and successively for indium nitride \cite{Pecz2020}. Besides III-nitrides, this process can be applied to oxides \cite{Turker2022} and metals \cite{Briggs2020, Rajabpour2021}. For group III metals, this is not the only method. In particular, atomically thin gallium layers (gallenene) have been obtained by intercalation of gallium in graphene, starting from the molten metal \cite{Kochat2018, Wundrack2021}. 
\\ \indent
These techniques allow for the fabrication of large-area, high quality, two-dimensional sheets of non-exfoliable materials. Additionally, the materials are protected from environmental contamination and oxidation by the graphene capping layer. Moreover, the process is easy to scale up to the wafer size. Therefore, it opens avenues for the development of new nanoelectronic devices. 
Furthermore, this approach is useful for investigating exotic properties of the materials emerging at the 2D limit, such as superconductivity \cite{Li2023, Bersch2019}, spin polarizability \cite{Vera2024, Kajale2024}, topological states \cite{Schmitt2024}, and nonlinear optical response \cite{Nisi2020, Steves2020}.
\\ \indent
Despite the significant interest dedicated to this subject in recent years, there are still open questions on the atomic structure of the confined materials. Especially the atomic arrangement of ultra-thin gallium is not well understood. Owing to the weak Ga-Ga bonds, gallium exhibits characteristics such as a melting point close to room temperature and a rich phase diagram comprising many crystalline phases. The stable phase under ambient conditions is \text{$\alpha$}-gallium \cite{Sharma1962}. 
The most investigated forms of gallenene are a100-gallenene \cite{Tao2018, Kochat2018, Nakhaee2019} and b010-gallenene \cite{Kochat2018, Nakhaee2019}. These, as suggested by their names, are derived from the (100) and (010) crystal planes of the orthorhombic lattice of bulk \text{$\alpha$}-gallium. The former is the analogous of graphene and exhibits a honeycomb lattice, while the latter is composed of a buckled hexagonal lattice. 
\text{$\beta$}-gallium is a metastable phase of gallium \cite{Bosio1981}. Its two-dimensional form has been experimentally observed \cite{Li2019}, and theory predicts that it can be achieved by relaxing a100-gallenene \cite{Anam2021}. Furthermore, theoretical studies predict several other stable gallenene allotropes. Examples are those emerging from the \text{$\gamma$} phase of Ga \cite{Gutierrez2020} and the variety of stochastically produced structures discussed in Reference~\cite{Kutana2022}. Finally, the centered tetragonal Ga(III) is one of the most elusive phases of Ga owing to the high pressure required for its formation \cite{Bosio1978}. Nevertheless, evidence of the formation of thin islands of Ga(III) exists \cite{Li2019}.
\\ \indent
The scope of this work is to shed light on the atomic configuration of 2D-Ga intercalated in EG and the emerging superstructures through surface-sensitive characterization techniques.
We have fabricated and characterized samples of two-dimensional gallium intercalated in epitaxial graphene. Via scanning tunneling microscopy (STM) and low-energy electron diffraction (LEED) measurements, we observed the appearance of unprecedented gallenene structures. The arrangement of gallium atoms has been analyzed to explain the emerging structures and compared to models. 

\begin{figure*}
\includegraphics[width=1\textwidth]{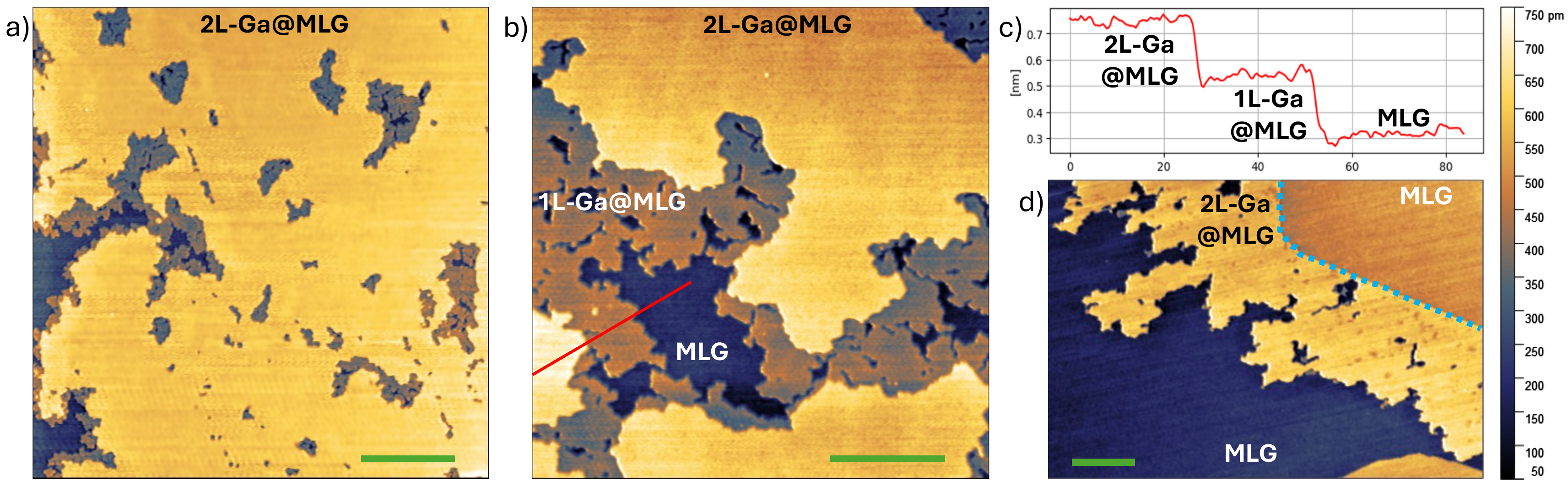}
\caption{\label{fig:a} \textbf{(a)} STM image (0.4~V, 0.8~nA) of 2D-gallium intercalated over large area. 2L-Ga@MLG is represented in yellow, pristine MLG in blue. \textbf{(b)} STM image (0.1~V, 0.7~nA) of an area in which both layers of an intercalated island are visible. The upper level (2L-Ga@MLG) is represented in yellow, the lower  (1L-Ga@MLG) in orange, and MLG in blue. \textbf{(c)} Height profile acquired along the red line in (b) which shows that both steps are about 0.2~nm high. \textbf{(d)} STM image (1.0~V, 1.0~nA) of an intercalated area originating from a SiC step edge (indicated by the light blue dashed line). Scale bars in (a), (b), and (d) are 100~nm, 50~nm, and 50~nm, respectively. Right: color scale for (a), (b), and (d).}
\end{figure*}

\section{Results}\label{sec:results}
\vskip-0.1cm
All 2D-Ga samples are fabricated starting from epitaxial graphene grown on SiC (refer to Section \ref{subsec:grahene growth}). Then, Ga is evaporated on the EG in a molecular beam epitaxy (MBE) chamber (see Section \ref{subsec:ga deposition}). Three samples were prepared with different deposition conditions. Finally, the intercalation of Ga atoms beneath the graphene sheet is achieved by thermal annealing under ultra-high vacuum (UHV). 
\vspace{-0.2cm}
\subsection{Two-dimensional Gallium Intercalation}\label{subsec:interc}
\begin{figure*}[h]
    \includegraphics[width=\textwidth]{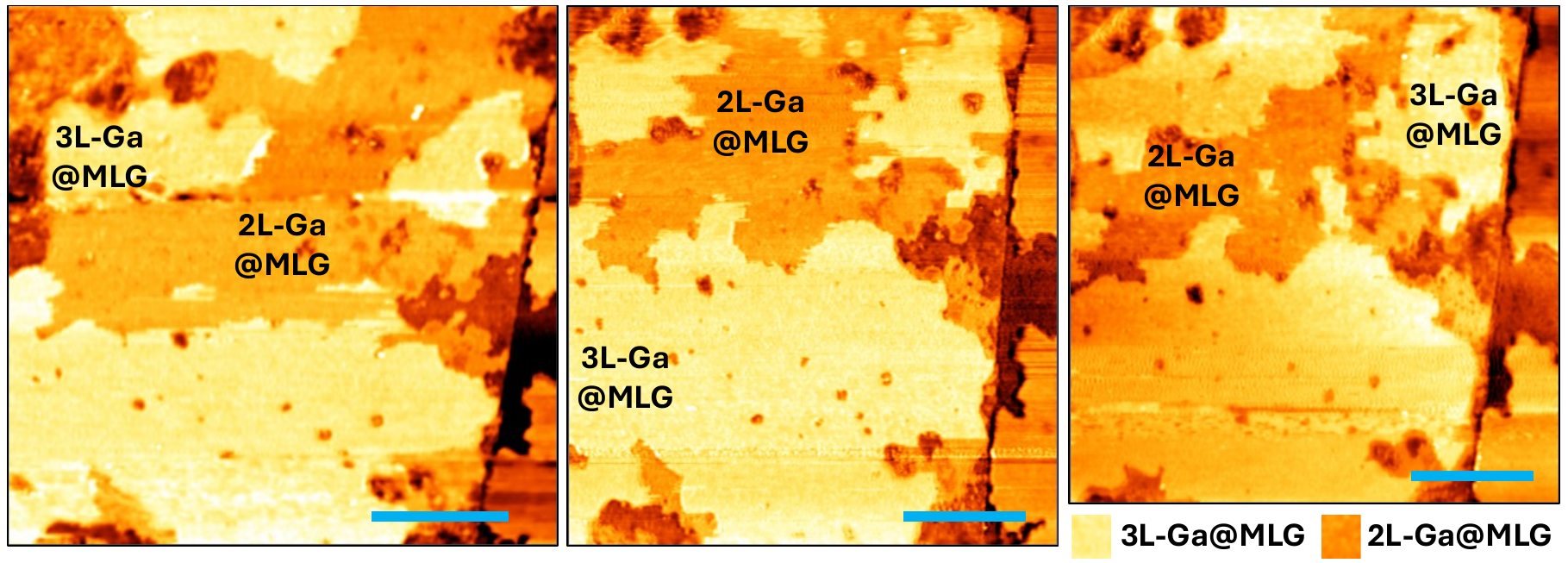}%
    \caption{STM images acquired at RT in succession on 3L-Ga@MLG on EG1 after the first annealing at 200~°C. The yellowish area, that is 3L-Ga, changes continuously shape after each scan. The scale bar indicates 50~nm in all images.} \label{SI:3L}
\end{figure*}

\vskip-0.1cm
During the deposition of Ga in the MBE chamber, the pristine EG samples are maintained at a constant temperature of 500~°C. 
Right after Ga evaporation, intercalation is almost absent, as shown in Figure~\ref{SI:afm}, and the majority of Ga remains on top of the EG in the form of large droplets.
This occurs because the deposition time scale is $5$ to $10$ seconds (refer to Table~S\ref{tab:MBE}), whereas the intercalation and diffusion of Ga atoms beneath the EG requires longer time to occur.
Therefore, thermal annealing has been performed after the deposition to achieve intercalation.
\\ \indent
On sample EG1, mainly composed of monolayer graphene (MLG), about one monolayer of Ga atoms was deposited. After the first annealing step at 200~°C for 12 hours under UHV, STM data shown in Figure~\ref{fig:a}(a) reveal the presence of $\mathrm{\mu m ^2}$-sized islands. Above these islands, we measured the graphene lattice with atomic resolution (see Figure~\ref{SI:graphene}). This confirms that Ga atoms have intercalated below the graphene sheet. The intercalated areas cover approximately 50\% of the total area of the sample, and the intercalated islands are uniformly distributed. 
Figure~\ref{fig:a}(b) shows that the islands exhibit two distinct height levels. The step height between these levels was measured via STM and is shown in the height profile in Figure~\ref{fig:a}(c). The step height between  MLG and the first level is $0.201\pm0.011~\mathrm{nm}$, and a similar height of $0.203\pm0.010~\mathrm{nm}$ is measured between the first and second level. A height of about 0.2~nm is compatible with the height of a single layer of Ga atoms \cite{Fohn2024}. Therefore, we attribute the first level to intercalated monolayer gallium (1L-Ga@MLG), and the second level to the intercalation of two layers of Ga (2L-Ga@MLG).
\\ \indent
The majority of the intercalated areas is composed of 2L-Ga@MLG. Small patches of 1L-Ga@MLG are visible at the edges of the 2L-Ga@MLG areas. Rarely, we observed regions of three-layer intercalated Ga (3L-Ga@MLG) (cf. Figure~\ref{SI:3L}). In these regions, the motion of the STM tip induces significant atomic mobility within the 3L-Ga@MLG islands, resulting in continuous changes to their morphology (as shown in Figure~\ref{SI:3L}). A previous study observed that the third layer exhibits a different atomic arrangement with longer bond lengths \cite{Briggs2020}. This is consistent with our observation that 3L-Ga@MLG islands appear to be unstable, differently from the first two layers.  
\\ \indent
\begin{figure*}[ht]
\includegraphics[width=1\textwidth]{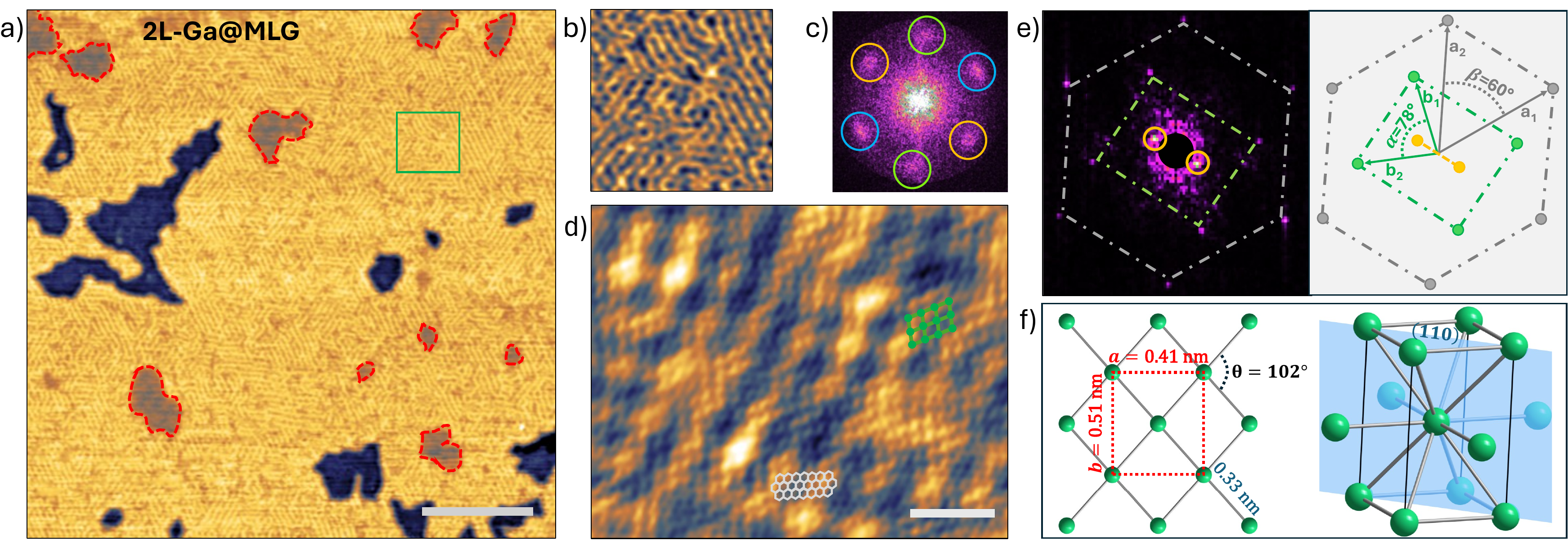}
\caption{\label{fig:b} \textbf{(a)} STM image (3.0~V, 0.8~nA) of 2L-Ga@MLG in which the striped pattern is imaged. 2L*-Ga@MLG regions are highlighted by the red dashed lines. \textbf{(b)} Close-up of (a) (corresponding to the green square) for a better visualization of the striped pattern. \textbf{(c)} FFT of (a). The observed hexagonal pattern is given by three pairs of spots, one per each orientation of the stripes (each orientation is highlighted by a different color). \textbf{(d)} Atomically resolved STM image (0.24~V, 0.6~nA) in which the stripes are visible, the graphene lattice, and the centered rectangular lattice of Ga(III). Sketches of the graphene and Ga lattices are superimposed to the image. \textbf{(e)} Left: FFT of (d). Graphene spots are highlighted by the grey hexagon and spots from gallium by the green rectangle. The spots from the striped moiré (only one direction is visible here) are circled in orange. Right: scheme of the spots in which unit vectors of Ga (b$_1$, b$_2$) and graphene (a$_1$, a$_2$) are indicated as well as the angle between them. \textbf{(f)} Left: schematic representation of the observed centered rectangular lattice of Ga(III). Right: unit cell of bulk Ga(III) where the (110) plane is highlighted in blue. Scale bars in (a) and (d) are 20~nm and 2~nm, respectively.}
\end{figure*}
Figure~\ref{fig:a}(d) shows an intercalated area originating at a SiC step edge (highlighted by a light blue dashed line). This suggests that discontinuities in the graphene sheet, such as the step edges of the substrate, promote the intercalation of Ga. As Ga atoms intercalate in graphene, they diffuse for hundreds of nanometers, creating the intercalated islands. Our observation is consistent with previous results on intercalated graphene \cite{Fiori2017}. Besides substrate step edges, defects within the graphene are also known to be major sites for intercalation \cite{Nayir2022}. Despite the high quality of the EG samples utilized here, defects are present. They play a fundamental role in the intercalation of Ga at locations far from SiC step edges. Indeed, as shown in Figure~\ref{fig:a}(a), large intercalated islands are observed even several micrometers away from the step edges.

\subsection{Intercalated Gallenene Structures}\label{subsec:atom}
\vskip-0.1cm
Upon closer examination of 2L-Ga@MLG, we observe the presence of striped domains, as shown in Figure~\ref{fig:b}(a).  For a better visualization of the stripes, Figure~\ref{fig:b}(b) shows a close-up of Figure~\ref{fig:b}(a). The stripes are evenly spaced by ($1.2\pm0.1$)~nm. They are not randomly oriented but follow a three-fold rotational symmetry. The fast Fourier transform (FFT) of Figure~\ref{fig:b}(a), shown in Figure~\ref{fig:b}(c), captures the symmetry of the structure. The hexagonal pattern is composed of the superposition of three pairs of spots, with one pair for each rotational stripe domain.
\\ \indent
Figure~\ref{fig:b}(d) shows an atomically resolved STM image of a striped domain. Here, the stripes are aligned along just one of the three possible orientations. The FFT of Figure~\ref{fig:b}(d), shown in Figure~\ref{fig:b}(e), reveals only two spots (highlighted by orange circles) which correspond to the orientation of this striped phase. The atomically resolved STM image (Figure~\ref{fig:b}(d)) and its FFT (Figure~\ref{fig:b}(e)) reveal that the stripes have the same orientation as the graphene lattice and correspond to a $5 \times 1$ structure with respect to graphene. 
\\ \indent
Interestingly, in addition to the striped phase and the graphene lattice (sketched in gray in Figure~\ref{fig:b}(d)), an additional periodicity is observed. In fact, the FFT reveals a rectangular pattern (highlighted in green). The rectangle is formed by the first-order spots of a centered rectangular lattice. Its base vectors (b$_1$, b$_2$) have equal lengths and include an angle of $\alpha$ = 78°. A schematic representation of the spots and unit vectors is shown in Figure~\ref{fig:b}(e). In real space, the unit vectors have a length of 0.33~nm and include an angle of 102°.
\\ \indent
The emerging rectangular lattice is attributed to the presence of gallium in the Ga(III) phase, because the Ga(III) phase exhibits a centered tetragonal lattice \cite{Bosio1978}. Its (110) crystal plane (depicted in Figure~\ref{fig:b}(f)) consists of a centered rectangular lattice with sides $a = 0.40$~nm and $b = 0.46$~nm. These values agree well with the experimental data $\tilde{a}=0.41\pm0.02$~nm and $\tilde{b}=0.51\pm 0.03$~nm. Moreover, the measured step height between MLG and 2L-Ga@MLG of $0.40\pm0.02$~nm is consistent with twice the distance between (110) crystal planes in bulk Ga(III) of 0.198~nm. 
\\ \indent
Under ambient conditions, it is not expected that Ga(III) will form \cite{Bosio1978}. 
Therefore, our observation of this phase of Ga suggests that the stabilization of Ga(III) is mediated by the interaction with SiC and graphene. 
Indeed, the rectangular lattice of Ga(III) is not randomly oriented. As shown in Figure~\ref{fig:b}(e), its major axis is aligned with one side of the hexagonal graphene ring, and as such, three possible orientations of the Ga(III) phase are possible.

\begin{figure*}[!t]
\vskip-0.05cm\includegraphics[width=1\textwidth]{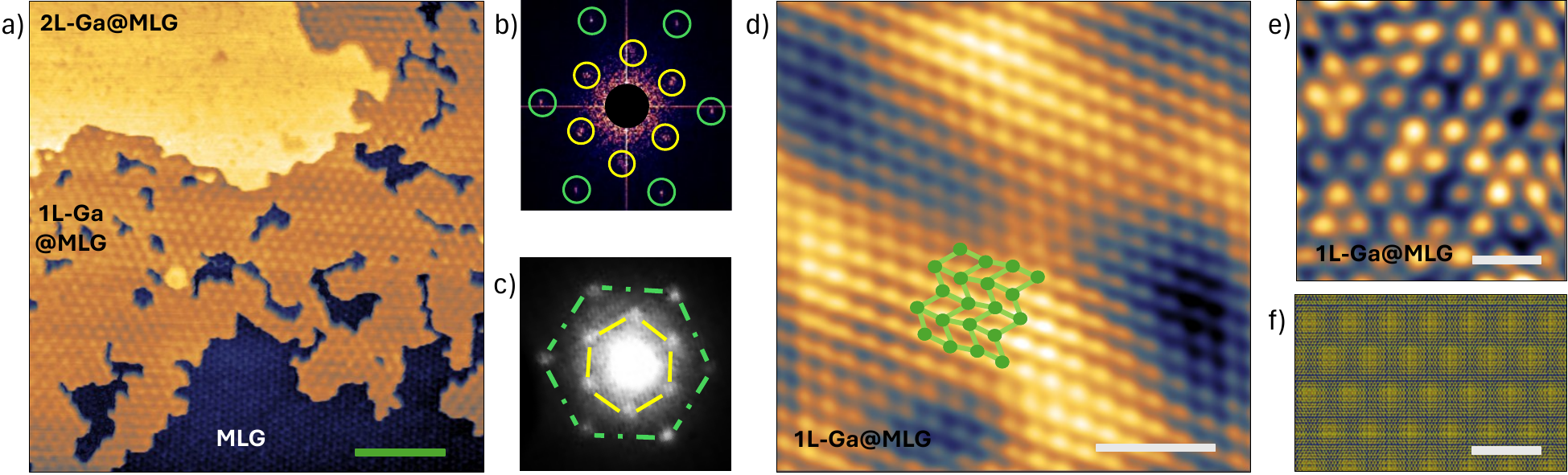}
\caption{\label{fig:c} \textbf{(a)} STM measurement (0.4~V, 0.8~nA) in a region mainly of 1L-Ga@MLG (represented in orange). Both the SiC-$6 \times 6$ moiré of pristine MLG and the graphene-$12 \times 12$ moiré on 1L-Ga@MLG are observed. \textbf{(b)} FFT of (a). The SiC-$6\times6$ and the graphene-$12 \times 12$ spots are highlighted in green and yellow, respectively. \textbf{(c)} LEED measurement performed on the same sample. The SiC-$6 \times 6$ and the graphene-$12 \times 12$ spots are highlighted with the same colors as in (b). \textbf{(d)} STM image ($-0.55$~V, $-0.31$~nA) acquired on 1L-Ga@MLG. A sketch of b010-gallenene is superimposed onto the measured lattice. \textbf{(e)} STM image (0.46~V, 0.12~nA) of the graphene-$12 \times 12$ moiré on 1L-Ga@MLG. \textbf{(f)} Simulated moiré given by the overlap of b010-Ga and graphene. The scale bars in (a), (d), (e), and (f) are 20~nm, 1~nm, 5~nm, and 5~nm, respectively.}
\end{figure*}

\begin{figure*}
    \includegraphics[width=0.5\textwidth]{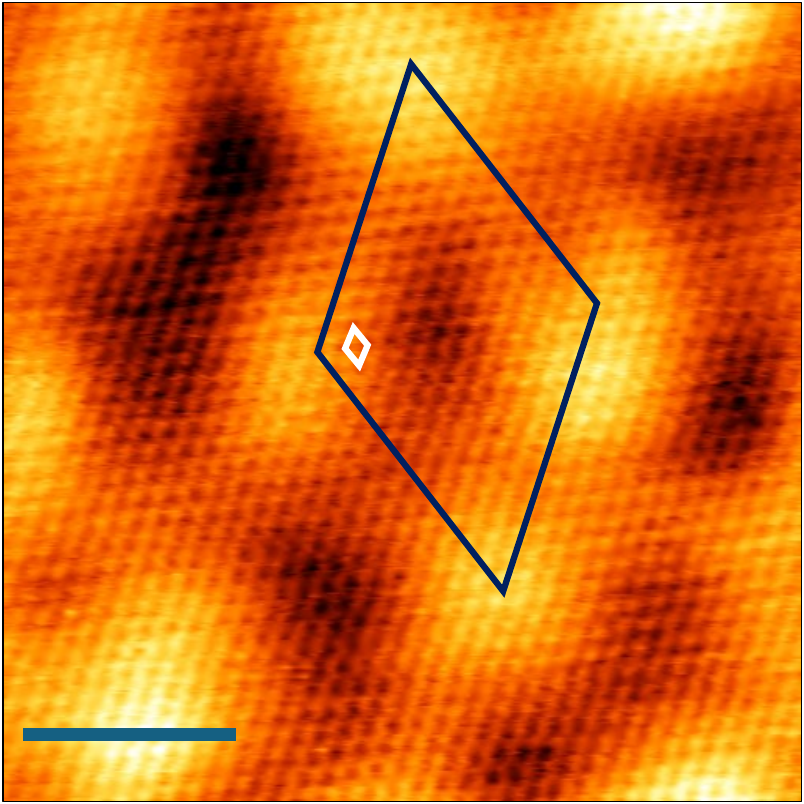}%
    \caption{STM image of 1L-Ga@MLG which shows the graphene lattice with atomic resolution. The unit cell of graphene is depicted in white and the 12$\times$12 cell in dark blue. The image is acquired in the same region as Fig. 3(d). Here the bias applied to the tip is $+0.4$ V instead of $-0.55$ V of Fig. 3(d). The scale bar indicates 2~nm.} \label{SI:bias}
\end{figure*}

\begin{figure*}
    \includegraphics[width=0.5\textwidth]{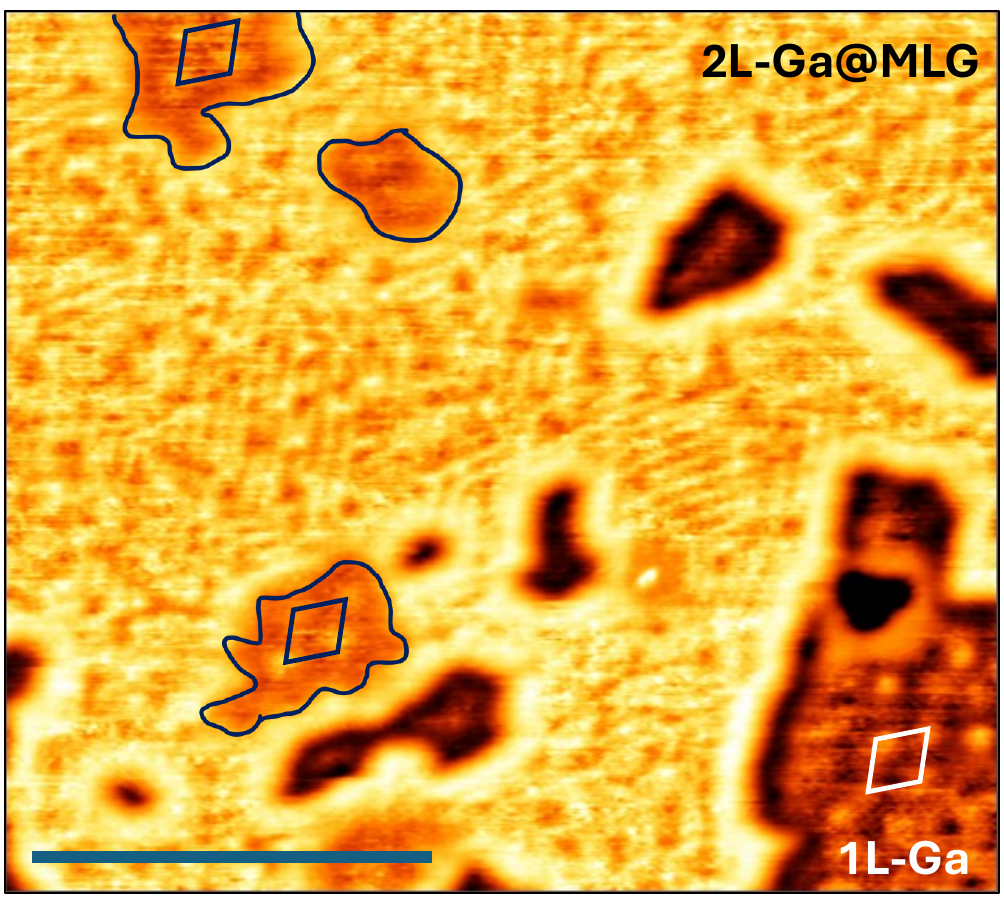}%
    \caption{STM image which shows regions of 2L*-Ga@MLG surrounded by 2L-Ga@MLG. 2L*-Ga@MLG areas are indicated by solid blue lines. These regions exhibit a hexagonal moiré which corresponds to the Gr-12$\times$12 moiré observed on 1L-Ga@MLG (bottom right of the image). The unit cell of the 12$\times$12 moiré is superimposed onto 2L*-Ga@MLG (blue) and 1L-Ga@MLG (white). The scale bar indicates 20~nm.} \label{SI:buche}
\end{figure*}

Moreover, the orientations of the rectangular lattice and the striped pattern are strictly related. The FFT in Figure~\ref{fig:b}(e) shows that the spots of the striped phase are parallel to the major axis of the rectangular pattern. This, in turn, is parallel to one of the graphene sides. Therefore, we conclude that the striped pattern is a moiré pattern emerging from the superposition of the honeycomb lattice of graphene and the rectangular lattice of Ga(III). The three possible orientations of the striped phase are thus determined by differently oriented Ga(III) grains.
\\ \indent
Furthermore, the periodicity $\lambda$ of the moiré pattern can be calculated as \cite{Liu2018}:
\[\lambda = \frac{b}{\sqrt{\delta^2+2(1+\delta)(1-cos(\theta/2))}} \;,\]
where $b = 0.51$~nm and $\theta$~=~102° are the parameters of the underlying rectangular Ga(III) lattice (as depicted in Figure~\ref{fig:b}(f)), and $\delta=b/(\text{g}\sqrt{3})-1$ with $g = 0.246$~nm the lattice parameter of the graphene overlayer. 
The formula predicts $\lambda=1.26$~nm, consistent with the measured value of $1.2\pm0.1$~nm.
\\ \indent
Unlike 2L-Ga@MLG, 1L-Ga@MLG exhibits a hexagonal moiré pattern. Figure~\ref{fig:c}(a) shows an STM image that exhibits the moiré on 1L-Ga@MLG.
This superstructure has a periodicity of $2.9 \pm 0.1$~nm and the same orientation as the graphene lattice. Therefore, it is a $12 \times 12$ modulation with respect to the graphene unit cell. The FFT of the STM image, which is shown in Figure~\ref{fig:c}(b), captures the emerging $12 \times 12$ moiré pattern (highlighted in yellow), as well as the SiC-$6 \times 6$ moiré pattern of non-intercalated EG (highlighted in green) \cite{Goler2013}. A LEED measurement performed on the same sample is shown in Figure~\ref{fig:c}(c). The diffraction pattern is in perfect agreement with the FFT result. This further confirms the presence of an ordered $12 \times 12$ moiré pattern.
\\ \indent
Imaging 1L-Ga@MLG with atomic resolution at a bias of $+0.4$~V, the graphene lattice was observed (see Figure~\ref{SI:bias}). Interestingly, as shown in Figure~\ref{fig:c}(d), at a larger and negative bias of $-0.55$~V, the STM image shows a different hexagonal lattice, from which a nearest-neighbor distance of $0.227\pm0.05$~nm is extracted. This structure is compatible with the planar projection of the buckled hexagonal lattice of b010-gallenene \cite{Kochat2018}. 
The observation of b010-gallenene is in fact consistent with the appearance of the $12 \times 12$ moiré pattern, which emerges from the superposition of gallenene and graphene. A $12 \times 12$ pattern was recently reported also on Pt-intercalated graphene \cite{Ferbel2025}. By comparing high-resolution STM data of the moiré, shown in Figure~\ref{fig:c}(e), and the simulated overlap between b010-gallenene and graphene, shown in Figure~\ref{fig:c}(f), the match is evident. In addition, the moiré period can be computed to be 2.93~nm (using the formula provided in Reference~\cite{Yankowitz2012}), in good agreement with the STM measurements.
\\ \indent
It should be noted that inside the 2L-Ga@MLG phase there are small inclusions without the striped pattern (2L*-Ga@MLG, indicated by red dashed lines in Figure~\ref{fig:b}(a)). 2L*-Ga@MLG does not exhibit the Ga(III) rectangular lattice. Instead, it shows a hexagonal pattern analogous to the $12 \times 12$ moiré of 1L-Ga@MLG (refer to Figure~\ref{SI:buche}). This different arrangement influences not only the surface modulation, but also the height of the layers. In fact, the surface of 2L*-Ga@MLG appears $\sim 50$~pm below the surface of 2L-Ga@MLG. This indicates a different arrangement of the intercalated Ga atoms, as will be discussed in Section \ref{subsec:model}. However, despite this slight variation in the height, 2L*-Ga@MLG is still compatible with two layers of Ga intercalated in graphene.
\\ \indent
In summary, we observed the coexistence of different phases of gallium in the same MLG sample, under the same conditions. This is consistent with the predicted small differences in free energy among the different phases of confined Ga \cite{Li2019}.

\begin{figure}[t]
\includegraphics[width=0.45\textwidth]{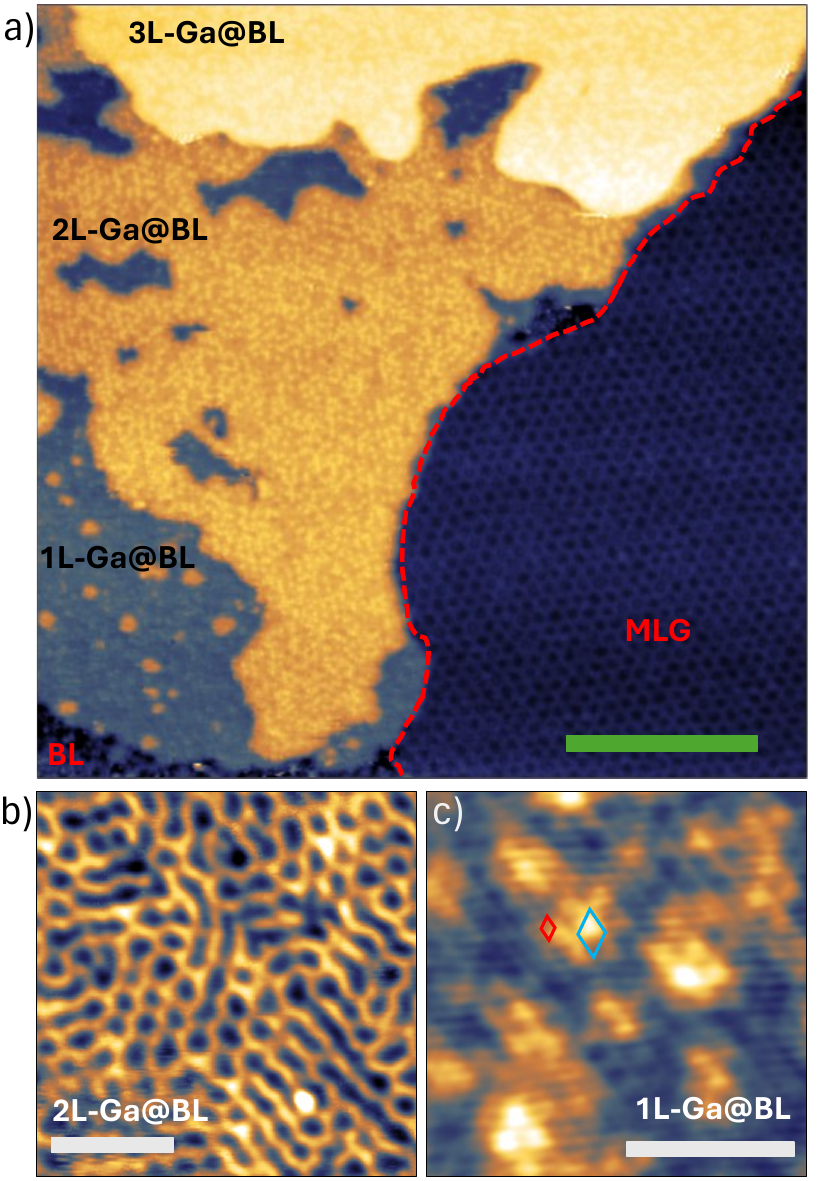}
\caption{\label{fig:d} \textbf{(a)} STM image ($-2.3$~V, $-0.62$~nA) of Ga intercalated on BL. The red dashed line indicates the edge between BL (on the left) and MLG (on the right). All three intercalated layers of Ga are imaged. \textbf{(b)} High magnification STM image (2.0~V, 0.5~nA) of the striped pattern observed on 2L-Ga@BL. \textbf{(c)} STM image ($-0.64$~V, $-0.24$~nA) acquired on 1L-Ga@BL exhibiting both the graphene lattice (unit cell represented in red) and the graphene-$2 \times 2$ reconstruction due to Ga (unit cell represented in light blue). The scale bars in (a), (b), and (c) indicate 20~nm, 5~nm, and 2.5~nm, respectively.}
\end{figure}

\subsection{Intercalation under the Buffer Layer} \label{subsec:buffer}
\begin{figure}[h!]
    \includegraphics[width=\textwidth]{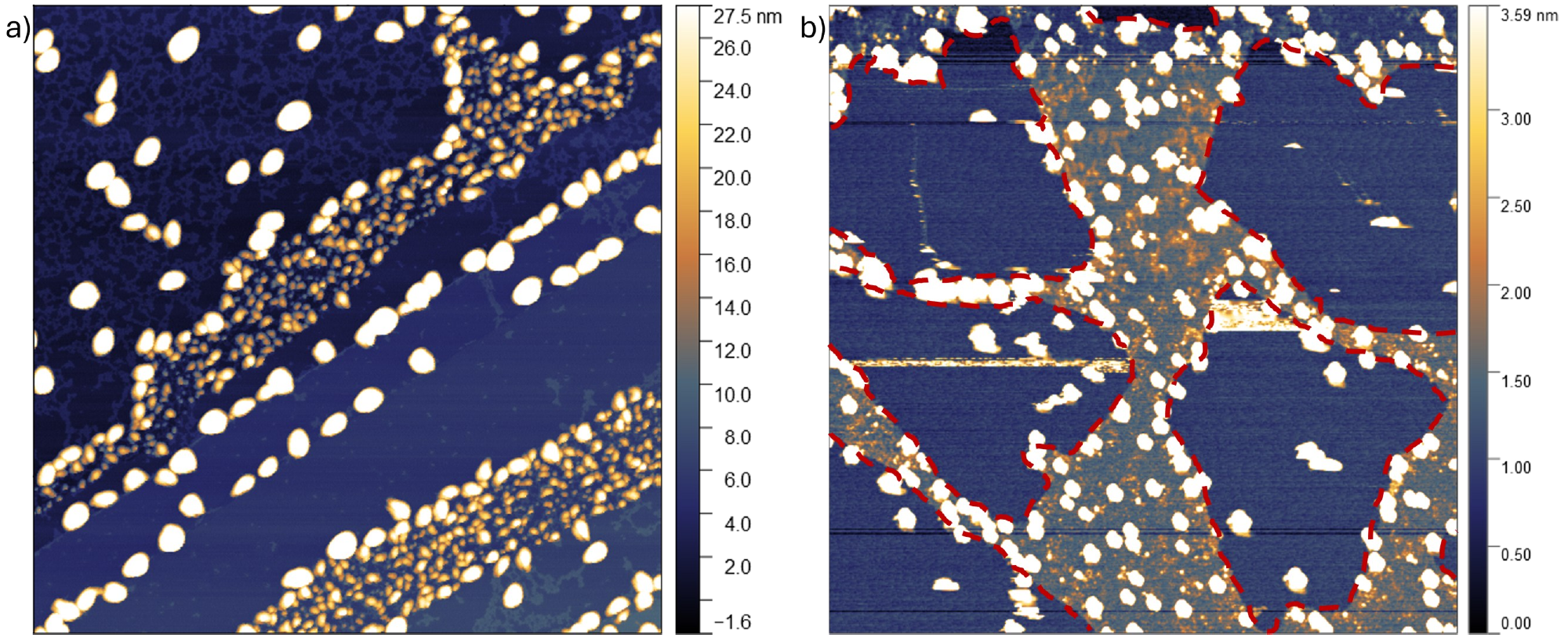}%
    \caption{2$\times$2 \text{$\mu m^2$} AFM images of (a) mainly MLG sample (EG1) and (b) mainly BL sample (EG2). In (b) the boundaries between MLG and BL are highlighted by the red dashed lines. In both samples right after the Ga deposition, BL is more decorated by Ga droplets (bright spots in the images). } \label{SI:afm}
\end{figure}

\begin{figure*}
    \includegraphics[width=\textwidth]{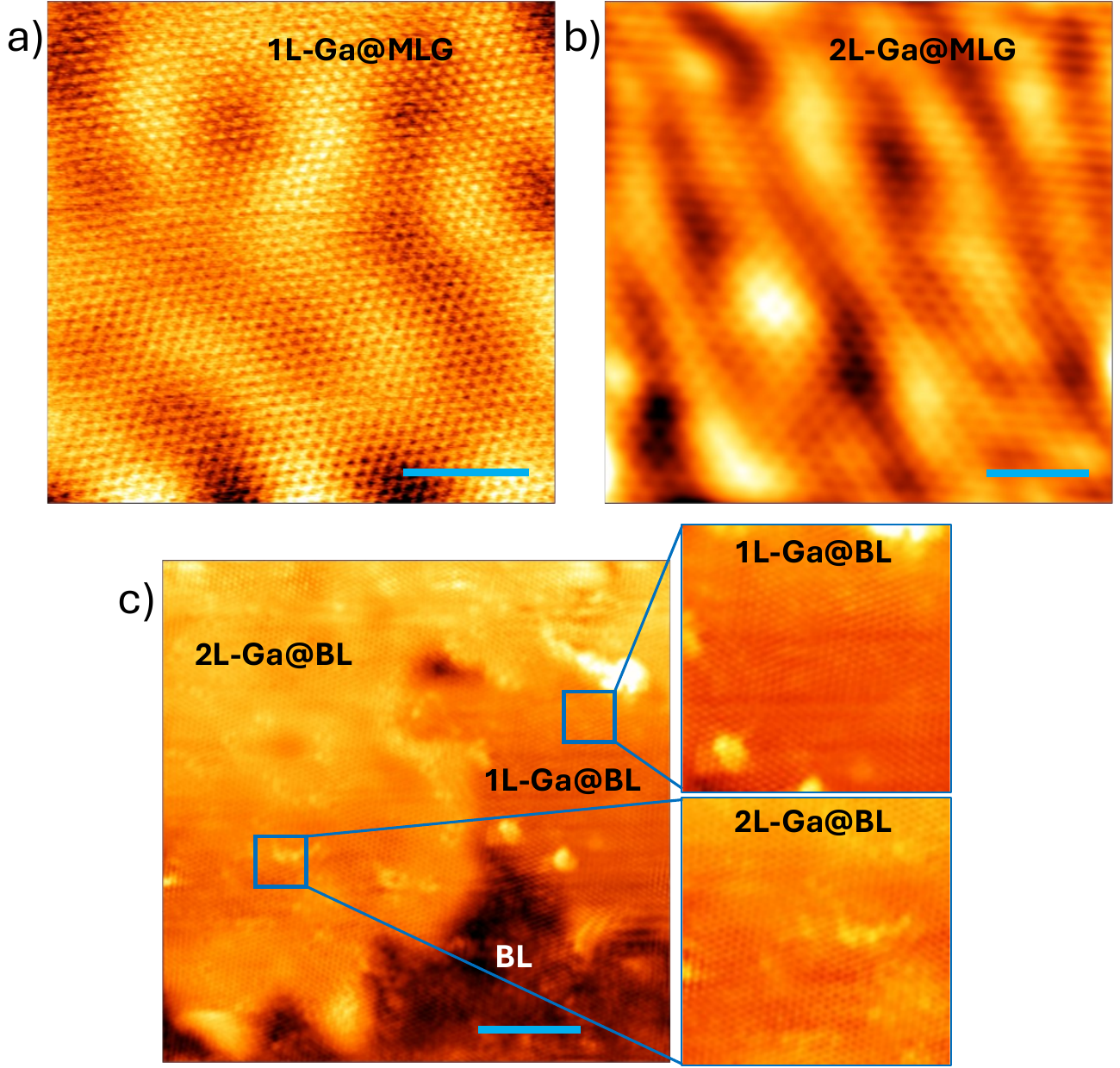}%
    \caption{Atomically resolved STM images showing the graphene lattice above (a) 1L-Ga@MLG, (b) 2L-Ga@MLG, (c) 1L- and 2L-Ga@BL. Close-ups of (c) are shown in the insets for a better visualization of the graphene lattice. The scale bars indicate 2~nm, 1~nm, and 4~nm, respectively. All images have been subjected to FFT filtering.} \label{SI:graphene}
\end{figure*}

\vspace{-0.1cm}
Besides the deposition of Ga on MLG, we explored the deposition of Ga on samples mainly composed of buffer layer (BL) (EG2 and EG3; see Section \ref{subsec:ga deposition}).
\\ \indent
Figure~\ref{fig:d}(a) shows an STM measurement of a BL region after deposition of Ga (acquired on the EG3 sample). Besides pristine BL regions that show the characteristic SiC-$6 \times 6$ superstructure \cite{Goler2013}, there are islands, not observed in the pristine sample, which emerge from the BL and do not show the SiC-$6 \times 6$ reconstruction. High-resolution STM measurements resolved the graphene lattice above all investigated regions (refer to Figure~\ref{SI:graphene}). This suggests intercalation of Ga in BL and the formation of quasi-freestanding MLG (QFMLG) \cite{Riedl2009, Goler2013}.
\\ \indent
The intercalated islands consist of three levels. The height differences between these levels extracted from STM data are $0.15\pm0.02~\mathrm{nm}$, $0.20\pm0.02~\mathrm{nm}$, and $0.16\pm0.01~\mathrm{nm}$ between BL and 1L-Ga@BL, from 1L-Ga@BL to 2L-Ga@BL, and from 2L-Ga@BL to 3L-Ga@BL, respectively. 
Therefore, similarly to MLG, each level is given by a single layer of gallium atoms. Here, gallium mainly intercalates as a trilayer (3L-Ga@BL). Areas of bilayer gallium (2L-Ga@BL) are observed around the trilayer islands. Rarely, small areas of monolayer gallium (1L-Ga@BL) are observed around 2L-Ga@BL. 
\\ \indent
Measurements performed on 2L-Ga@BL revealed the presence of a surface modulation resembling the striped moiré observed in 2L-Ga@MLG. Figure~\ref{fig:d}(b) shows an STM measurement of the striped pattern on 2L-Ga@BL. Here, the pattern appears less ordered compared to that in 2L-Ga@MLG. We suggest that this is due to the higher number of defects in the BL compared to MLG. Still, both the distance between the features and their orientation are consistent with the striped moiré in 2L-Ga@MLG. Since the striped moiré is produced by the lattices of graphene and Ga(III), as discussed in Section \ref{subsec:atom}, this allows to conclude that the interaction of Ga with graphene and SiC allows for the stabilization of Ga(III) also in the intercalated BL. Moreover, the presence of the same striped superstructure in both BL and MLG samples implies that Ga(III) forms between the buffer layer and the SiC, as will be discussed in Section~\ref{subsec:model}.
\\ \indent
The surface roughness of 1L-Ga@BL and 3L-Ga@BL ($25 \pm 5$~pm) is substantially smaller than 2L-Ga@BL ($45 \pm 5$~pm). High-resolution STM measurements on 1L-Ga@BL and 3L-Ga@BL reveal the presence of a $2 \times 2$ reconstruction with respect to graphene, as reported in Figure~\ref{fig:d}(c). The $2 \times 2$ registry of Ga atoms results from the occupation of alternating hollow sites of the graphene lattice. This behavior has also been reported for K- and Rb-intercalated graphene \cite{Huempfner2023, Ferbel2025arxiv}.
\\ \indent
The intecalated areas in both samples EG2 and EG3 exhibit the same structures, besides the different coverage of Ga on the samples.
\\ \indent
The measurements performed on these samples (with 60\% BL and 40\% MLG) revealed that Ga is much more prone to adhere to BL. AFM measurements performed right after the Ga deposition show that BL was much more decorated by Ga droplets than MLG (as shown in Figure~\ref{SI:afm}). Moreover, STM measurements demonstrated that Ga intercalation preferentially occurred in BL areas.
This is ascribed to the higher density of defects and the larger surface roughness characteristic of the BL compared to MLG. Thus, BL offers a larger number of energetically favorable adsorption sites for gallium atoms \cite{Hu2021}. 
\vspace{-0.15cm}
\subsection{Thermal Stability of Gallenene Structures}\label{subsec:therm}
\begin{figure*}
    \includegraphics[width=\textwidth]{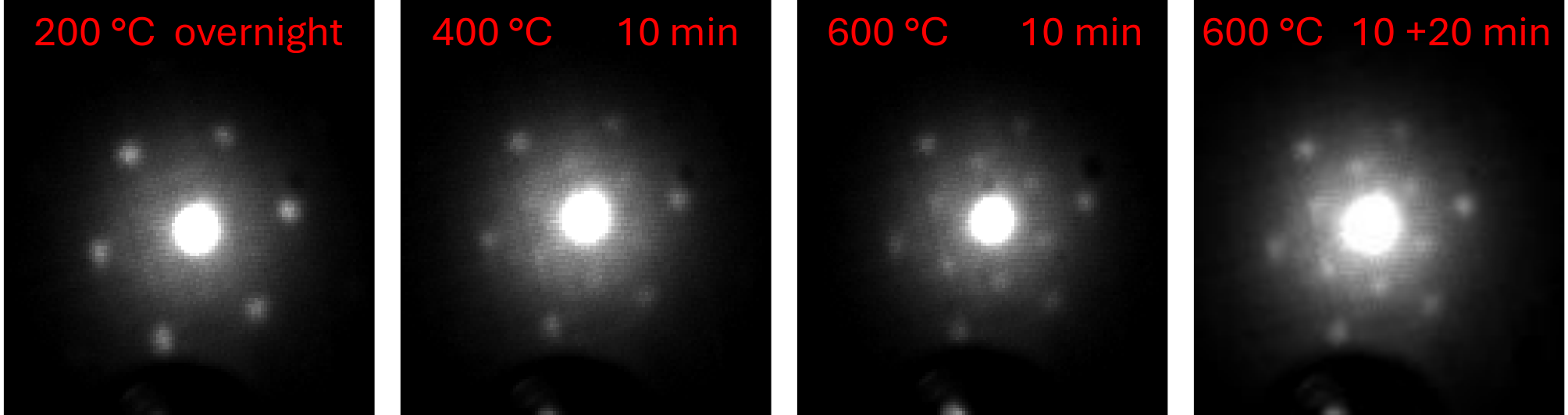}%
    \caption{Series of LEED patterns acquired on sample EG1 after every step of annealing (annealing procedure reported in each image). The larger hexagon, which corresponds to the 6$\times$6 moiré of pristine MLG, appears to remain constant with annealing. Differently, a smaller hexagonal pattern emerges upon annealing and becomes sharper with increasing annealing temperature. It corresponds to the 12$\times$12 moiré initially observed on 1L-Ga@MLG (and successively on 2L- and 3L-Ga@MLG).} \label{SI:LEED}
\end{figure*}

\begin{figure*}
    \includegraphics[width=\textwidth]{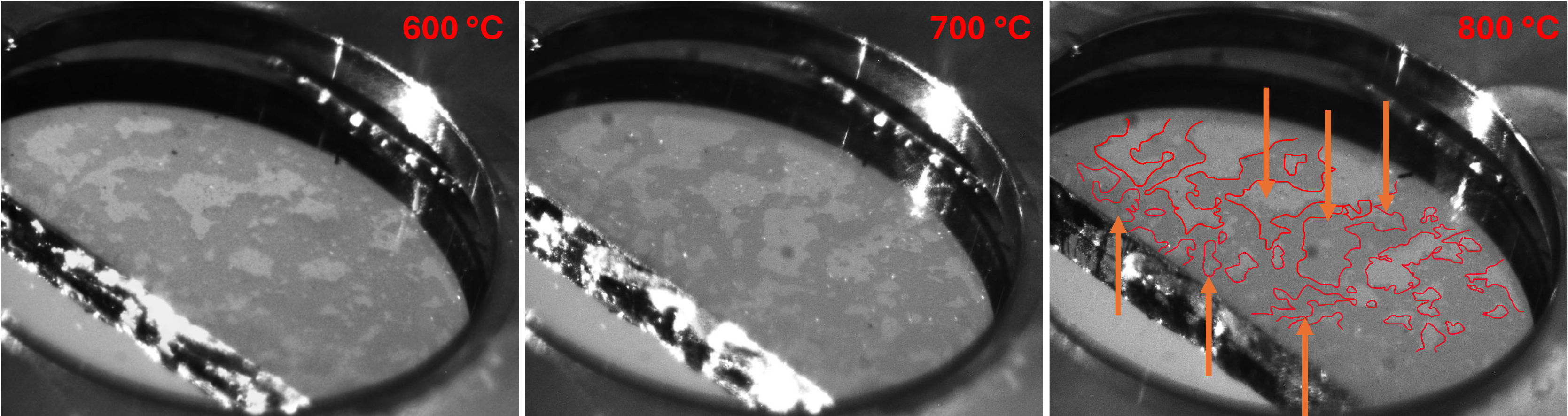}%
    \caption{Series of optical images (acquired with the video camera of the STM) of the MLG sample at different stages of annealing (annealing temperature reported in each image). From 600~°C (left) to 700~°C (center) the bright areas (regions rich in intercalated Ga) increase. Instead, upon annealing at 800~°C (right picture) the size of the bright areas is reduced. The contours of the bright areas of the central picture are superimposed onto the right picture to highlight the missing parts indicated by arrows.} \label{SI:optic}
\end{figure*}

\vspace{-0.1cm}
Results reported in Sections \ref{subsec:interc} to \ref{subsec:buffer} are related to samples annealed at 200~°C overnight under UHV. After this first thermal treatment, samples were subjected to annealing steps at gradually increasing temperatures up to 800~°C, while the evolution of the aforementioned structures was monitored.
\\ \indent
On the MLG sample (EG1), after Ga deposition and the first annealing at 200~°C, the most common configuration is 2L-Ga@MLG with the $5 \times 1$ striped moiré. Upon further annealing up to 400~°C, the $5 \times 1$ striped moiré is stable, whereas for temperatures higher than 400~°C, it gradually weakens, until it completely disappears at around 600~°C. In its place, a hexagonal moiré appears. This is due to the transition from 2L-Ga@MLG to 2L*-Ga@MLG. The emerging moiré has the same periodicity and dimensions of that observed on 1L-Ga@MLG, that is graphene-$12 \times 12$. The same moiré is also observed on 3L-Ga@MLG after annealing at 600~°C. The relaxation of all these layers to form the same moiré implies a significant increase in  surface area exhibiting the $12 \times 12$ moiré.
This is consistent with the LEED measurements performed after each annealing step. Figure~\ref{SI:LEED} shows the evolution of the LEED pattern with  thermal treatments. Initially, after the first annealing at 200~°C, the LEED spots of the graphene-$12 \times 12$ are faint, consistent with STM observations of the moiré only in small patches of 1L-Ga@MLG and in the rare regions of 2L*-Ga@MLG. The intensity of the $12 \times 12$ LEED moiré spots gradually increases with increasing annealing temperature, reaching a maximum upon annealing at 600~°C. This is in agreement with the STM observation of a gradual spreading of the hexagonal moiré of 2L*-Ga@MLG, in place of the striped moiré of 2L-Ga@MLG. 
Moreover, with successive annealing steps, the intercalated area increases from about 50\% to about 90\% upon annealing at 700~°C. This is due to the fact that, initially, part of the gallium remained in large droplets above the graphene sheet. Annealing facilitates the diffusion of Ga atoms from the droplets to the surrounding graphene, resulting in further intercalation.
\\ \indent
Finally, sample EG1 was heated to 800~°C. At this temperature, the de-intercalation and desorption of Ga from the surface is observed. Figure~\ref{SI:optic} shows a comparison between optical images (acquired from the video camera of the STM chamber) of the sample at different stages of annealing. The images confirm that the largest intercalation coverage is achieved upon 700~°C annealing. After annealing at 800~°C, a large fraction of the previously intercalated areas is missing. 
The de-intercalation process is known to be thermally activated in this system \cite{Niefind2023}. In our work, the process occurred at much higher temperature ($\sim800$~°C) compared to previous results ($\sim300$~°C) \cite{Niefind2023}. The greater thermal stability of our sample can be motivated by the different 2D-Ga fabrication procedures and the different defect densities of the utilized EG samples, which in Reference~\cite{Niefind2023} are treated with O$_2$/He plasma to increase the defectivity.
\\ \indent
Differently, the de-intercalation of Ga from BL is initiated at lower temperatures, closer to the $\sim$300~°C reported in Reference~\cite{Niefind2023}. Annealing a sample of Ga intercalated in BL at temperatures above 400~°C induces Ga diffusion to the surface. This results in the formation of Ga clusters above the BL, which are detected by STM measurements. 
In particular, the second and third layers of intercalated Ga are more prone to de-intercalate, whereas 1L-Ga@BL is still observed upon annealing at 600~°C. This is an indication of the strong interaction of Ga atoms with the SiC substrate.

\begin{figure}[b]
\includegraphics[width=0.5\textwidth]{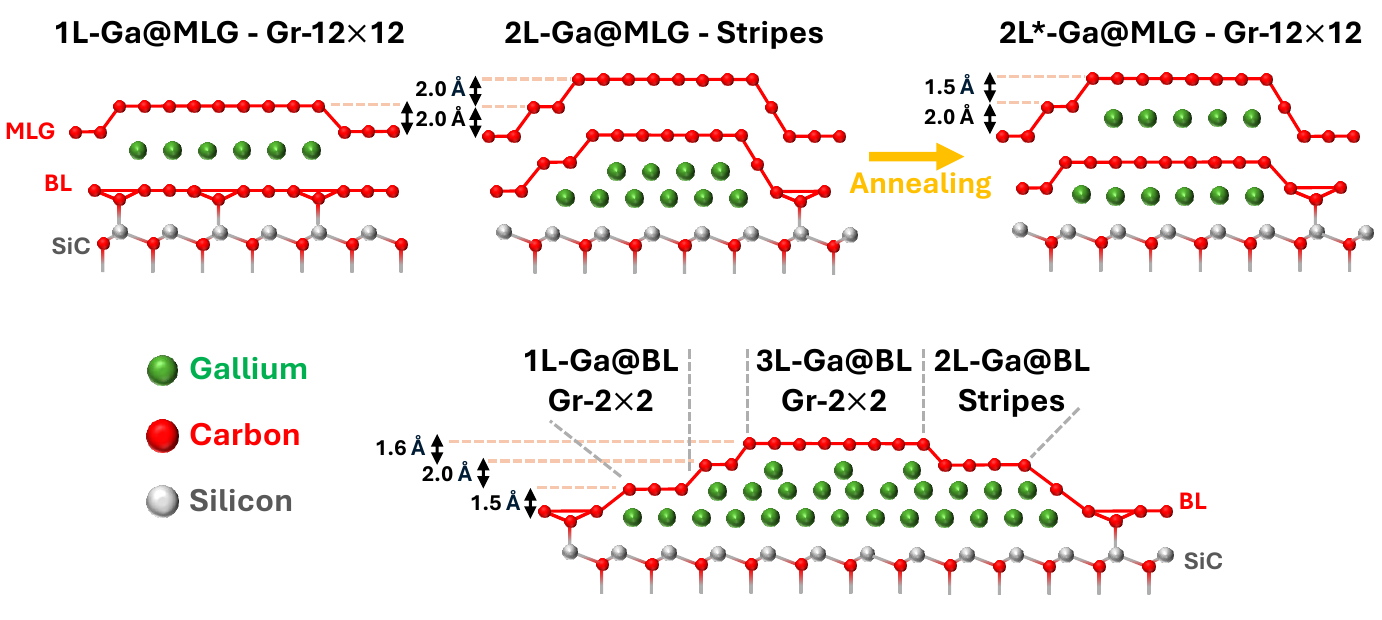}
\caption{\label{fig:e} Sketch of the proposed model for the intercalation of gallium in MLG (top) and BL (bottom).}
\end{figure}
\subsection{Modelization of the Intercalation Process}\label{subsec:model}
\vskip-0.1cm
The information of Sections \ref{subsec:atom} to \ref{subsec:therm} allowed us to create a model that represents the discussed structures. Figure~\ref{fig:e} shows a schematic representation of the model.
\\ \indent
The comparison of intercalated structures on MLG and BL allows to determine the relative position of the graphene sheets and the gallium layers. Islands of Ga intercalated in BL never exhibit a hexagonal $12 \times 12$ moiré. This suggests that the $12 \times 12$ superstructure observed in 1L-Ga@MLG in MLG corresponds to a Gr/Ga/BL/SiC configuration (see Figure~\ref{fig:e} top left), where the $12 \times 12$ moiré emerges from the overlap of the topmost graphene and b010-gallenene.
\\ \indent
The presence of a striped pattern on both 2L-Ga@MLG and 2L-Ga@BL suggests that the arrangement of Ga in  the striped pattern is the same in MLG and BL. This implies a (2L-)Gr/2L-Ga/SiC configuration ('2L-' between brackets refers to MLG; see Figure~\ref{fig:e} top center and bottom) in which the Ga(III) bilayer forms a centered rectangular lattice stabilized by the interaction with graphene and the SiC substrate.
\\ \indent
As discussed in Section \ref{subsec:therm}, the initially small regions of 2L*-Ga@MLG with a hexagonal moiré (cf. Figure~\ref{fig:b}(a)) spread and replace the striped regions with thermal annealing. The appearance of a $12 \times 12$ moiré occurs because of the superposition of a graphene layer to a gallenene layer. Therefore, the transition from a striped to a hexagonal superstructure is achieved by the thermally activated climbing of part of the Ga(III) atoms above the QFMLG to rearrange into monolayer gallenene. The climbing process is consistent with the theoretical prediction in Reference~\cite{Niefind2023}.  Figure~\ref{fig:e} (top-right) shows the Gr/Ga/Gr/Ga/SiC configuration obtained upon climbing of Ga atoms.
\\ \indent
Finally, the model consistently explains the fact that de-intercalation in the BL is achieved at lower temperatures. Initially, in both 2L-Ga@MLG and 2L-Ga@BL, Ga atoms are located between QFMLG and the SiC substrate. If energy is provided (by annealing), Ga tends to climb above the QFMLG layer. This results in the growth of $12 \times 12$ moiré domains in intercalated MLG, while in BL it results in the de-intercalation.
This suggests that the initial configuration of Ga(III) is stable, but  not the absolute minimum energy state of the system. Therefore, upon annealing, the system relaxes into the most favorable configuration, which is one layer of Ga bonded to the SiC substrate and one layer above QFMLG. 

\section{Conclusions}\label{sec:conclusion} 
\vskip-0.29cm
We have provided an extensive analysis of the variety of phases and atomic arrangements which are obtained in the graphene-Ga-SiC system. 
By STM measurements we analyzed the evolution and structure of Ga intercalation in graphene upon thermal treatments.
Our measurements reveal the appearance of novel superstructures in this system, including a hexagonal Gr-$12 \times 12$ moiré and a striped Gr-$5 \times 1$ pattern. Atomically resolved STM measurements confirm the presence of b010-gallenene. It appears to be the most stable form of gallenene, consistently with other experimental observations \cite{Kochat2018} and theory \cite{Nakhaee2019, Kutana2022}. Additionally, the rare high-pressure phase Ga(III) was observed, and its stability was investigated.
Based on our findings, we built a model to explain all observed atomic arrangements. Our modelization is consistent with  theoretical calculations on Ag- and Ga-intercalated graphene \cite{Niefind2023}.
\\ \indent
Unlike many of the reported works on  graphene/gallenene/SiC \cite{Briggs2020, Wundrack2021}, here we deposit Ga via MBE under UHV conditions, and the intercalation is gradually induced by successive thermal treatments under UHV conditions. This allows to evolve the system step by step and to select the conditions to obtain a desired configuration.
Therefore, the fabrication procedure adopted here provides a playground to investigate new properties of the system in various configurations. This can be fruitful in the investigation of the superconductivity  of the Ga(III) phase \cite{Li2019, Kutana2022}, in exploring the moiré physics of the system, or in understanding metal-to-insulator transitions in gallenene \cite{Bondarenko2023}.
\\ \indent
The method employed here allows for the fabrication of a tunable platform in terms of atomic configuration selectivity. Additionally, the graphene capping ensures the environmental stability. Moreover, this fabrication procedure is projected towards the realization of different materials, both mono-elemental (e.g., In, Al, Sn, and Pb) and compounds (in particular, III-V semiconductors) with potentially groundbreaking implications for the development of novel devices.

\section{Methods}\label{sec:method}
\vskip-0.2cm
\subsection{Graphene Growth}\label{subsec:grahene growth}
\begin{figure*}[h]
    \includegraphics[width=\textwidth]{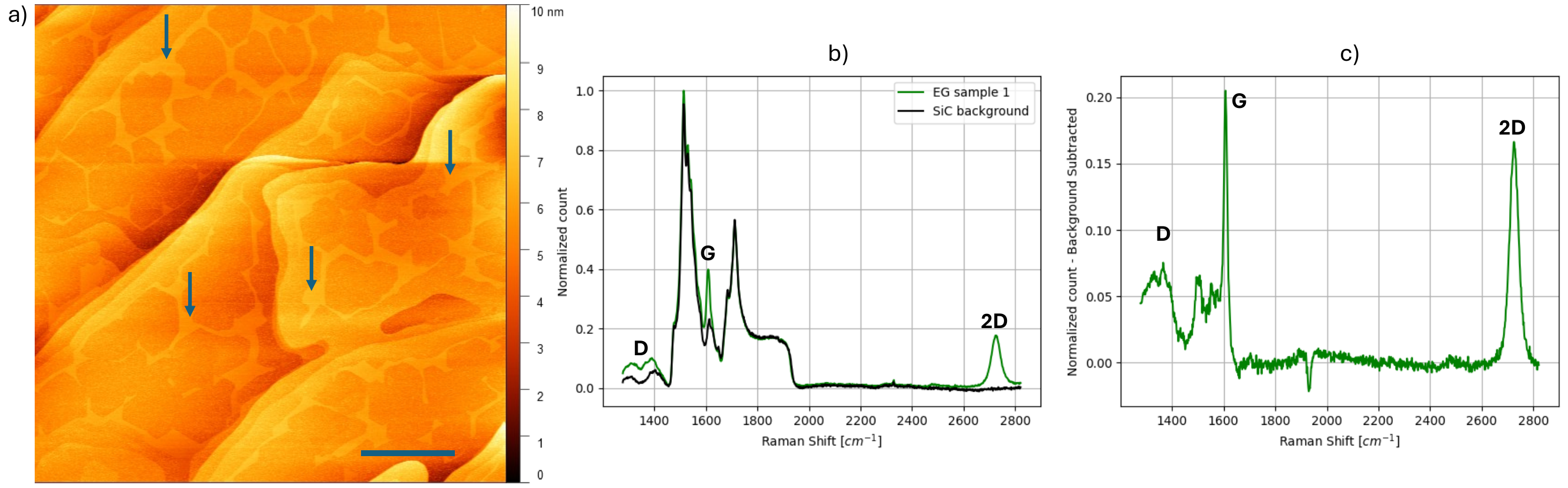}%
    \caption{(a) AFM image acquired on EG1 prior to deposition of gallium. AFM images have been utilized to estimate the MLG coverage, which is about 85\%. Arrows indicate some of the BL regions. The scale bar indicates 2~$\mu$m. (b) Raman spectrum of EG1 compared to the SiC Background signal. 2D and G clearly emerge from the background as highlighted in (c), where the SiC signal is subtracted.} \label{SI:S1}
\end{figure*}

\begin{figure*}[h]
    \includegraphics[width=\textwidth]{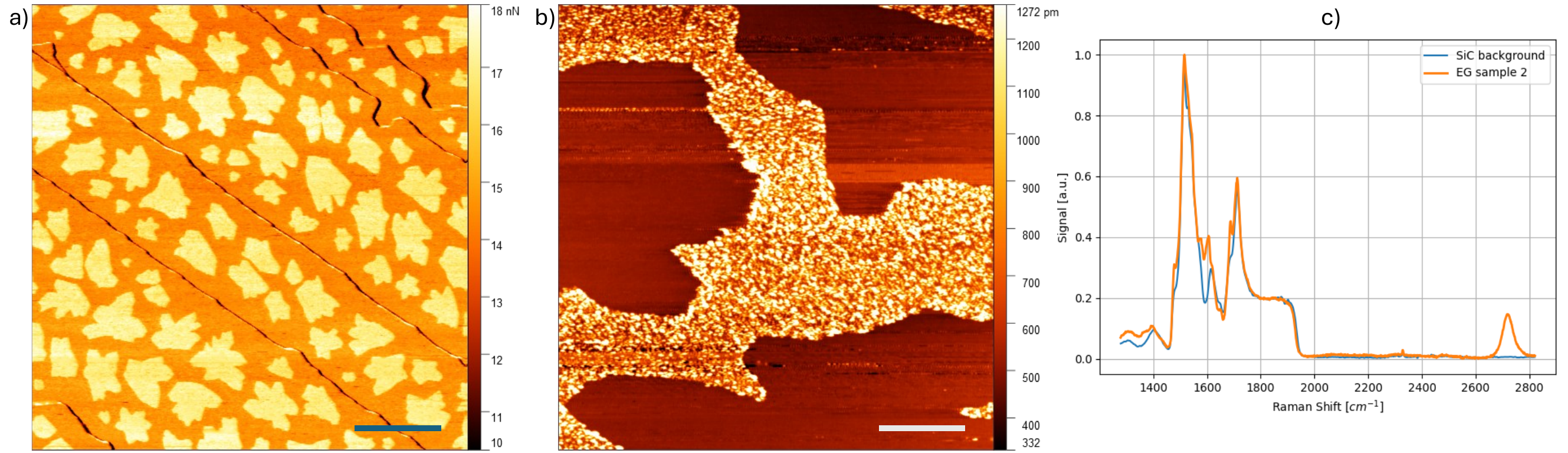}%
    \caption{(a) AFM image acquired on EG2 prior to deposition of gallium. For a higher contrast between MLG (brighter) and BL (darker) the image is not reported in height, but in adhesion scale. From AFM measurements, the MLG coverage is estimated to be about 40\%. (b) STM image of pristine EG2 which confirms that the sample is composed of islands of MLG (flat brown areas) surrounded by BL (bright and corrugated areas). (c) Raman spectrum of EG2 compared to the SiC background. Despite the lower coverage of MLG, the 2D peak is evident, while the G peak is less pronounced compared to EG1.} \label{SI:S2}
\end{figure*}

\vskip-0.1cm
Graphene samples are epitaxially grown on the Si-face of 6H-SiC substrates. SiC wafers are hydrogen-etched prior to graphenization to reduce the surface roughness.
Graphene growth is achieved via thermal decomposition of SiC conducted under an Ar atmosphere in a BM-Aixtron reactor at a temperature of approximately 1300~°C.  
\\ \indent
Three samples were produced. EG1 is composed of 85\% MLG and 15\% BL, while EG2 and EG3 are composed of 60\% BL and 40\% MLG. For more details on the pristine sample characterization, we refer to Figure \ref{SI:S1} and \ref{SI:S2}.

\subsection{Gallium Deposition}\label{subsec:ga deposition}
Pristine EG samples are loaded into an MBE system. Prior to gallium deposition, the samples are degassed under UHV (base pressure $4\times10^{-9}$~mbar) for 30 minutes at 300~°C. After degassing, the samples are moved to the growth chamber of the MBE, where they are exposed to a flux of pure gallium while being kept at 500~°C (temperature measured using an IR-pyrometer). Different samples were subjected to different deposition conditions (see Table~S\ref{tab:MBE}).

\subsection{Gallium Intercalation}\label{subsec:ga intercalation}
After Ga deposition, samples are transferred in air to the UHV system. Here, they are subjected to a series of annealing steps at increasing temperatures under UHV conditions (base pressure $5 \times 10^{-11}$~mbar). Samples are mounted on the sample holders on top of a Si substrate. During annealing steps, samples are heated via a direct current flow through the Si substrate. The temperature is monitored using an IR-pyrometer and a thermocouple mounted on the sample holder. 

\subsection{Characterization Methods}\label{subsec:characterization}
The UHV system is equipped with a LEED and an STM. STM measurements were conducted at room temperature using a VT-UHV STM from RHK. LEED measurements were conducted using a BDL-600IR from OCI Vacuum Microengineering (spot size $\sim$500~µm). 
\\ \indent
AFM measurements were performed ex-situ utilizing a Dimension Icon from Bruker, operated in tapping mode. 
The Raman spectrometer used is an inVia confocal Raman microscope (equipped with a 532~nm laser, with spot diameter of 1~$\mu$m) from Renishaw. 

\bibliography{bib}

\newpage
\onecolumngrid
\center\large{\textbf{SUPPORTING INFORMATION\\
Novel structures of Gallenene intercalated in epitaxial Graphene}}
\vskip3cm
\section{Pristine Epitaxial Graphene Samples}

\clearpage

\newpage

\section{Gallenene Samples}

\begin{table}[h!]
    \centering
    \begin{tabular}{c|c|c|c}
         Sample & EG1 & EG2 & EG3\\ \hline
         Flux [atoms$/(\text{cm}^2$ s)] \; & \; $\sim 6 \cdot$10$^{15}$ \; & \; $\sim 3 \cdot$10$^{15}$ \; & \; $\sim 2 \cdot$10$^{15} \; $ \\ \hline
         Time [s] & 10 & 5 & 5\\
    \end{tabular}
    \caption{Gallium deposition conditions for the different samples. All samples were kept at 500~°C during the Ga exposure.}
    \label{tab:MBE}
\end{table}

\clearpage

\end{document}